# ZEUS: An Experimental Toolkit for Evaluating Congestion Control Algorithms in 5G Environments

Rohail Asim
*NYU Abu Dhabi*

Muhammad Khan
*NYU Abu Dhabi*

Luis Diez
*University of Cantabria*

Shiva Iyer
*NYU*

Ramon Aguero
*University of Cantabria*

Lakshmi Subramanian
*NYU*

Yasir Zaki
*NYU Abu Dhabi*

## Abstract

As global cellular networks converge to 5G, one question lingers: Are we ready for the 5G challenge? A growing concern surrounds how well do existing congestion control algorithms perform in diverse 5G networks. Given that 5G networks are not yet widely deployed, assessing the performance of existing congestion control algorithms in realistic 5G settings presents several challenges. Moreover, existing network simulation and emulation environments are also not ideally suited to address the unique challenges of 5G network environments. Therefore, building a simple and easily accessible platform becomes crucial to allow testing and comparison of congestion control algorithms under different testing conditions.

This paper makes two main contributions. First, we present *Zeus*, an open-source testbed that emulates 5G channels to evaluate congestion control algorithms in a repeatable and reproducible manner. Second, we assess and characterize ten of today's state-of-the-art congestion control algorithms under different 5G environments and show the difficulties of these solutions to achieve high performance under highly varying 5G channel conditions. In addition, we also utilize the recently proposed harm metric to characterize the detriment done by such algorithms to TCP Cubic cross traffic in 5G environments.

## 1 Introduction

It seems impossible to evade the hype around 5G, which is intended to create unprecedented user experiences, transform industries and enrich lives [29]. However, a transformational technology like 5G will take time to evolve and mature. The 5G roll-outs by major cellular operators [18], [61] today are following the pragmatic Non-StandAlone (NSA) mode, reusing the legacy 4G infrastructure to reduce cost. With 5G deployments, operators seek to achieve multi-Gbps wireless bit-rate for bandwidth-hungry applications such as 4K/8K Ultra High Definition (UHD) video and Virtual Reality (VR) transmission, ultra-Reliable and Low Latency Communication (uRLLC) for auto-driving, or telesurgery [21]. However, it is still early to tell how well 5G can meet its full potential. A question still unanswered is *"How far is 5G from its prospects, and what are the missing pieces of the 5G puzzle?"*. For instance, the 5G New Radio (NR) has advanced swiftly between the 3GPP standardized Release 15 [46] and Release 16 (finalized) [45], enabling an extension to higher carrier frequencies to meet the continuous need for more traffic and higher data rates. However, the transport layer has not evolved at the same pace to address the challenges accompanied by 5G NR. This is the case of 5G millimeter-wave (mmWave) access technologies, which refer to higher spectrum bands in the range of 30 $\text{GHz}$ and 300 $\text{GHz}$ and exhibit substantial variations in their transmission capacity over short time scales. Such high fluctuations have led to several MAC/PHY layer solutions [48, 53, 57, 69]. However, lower-layer network innovations result in different forms of packet delay and rate variations over short time scales, introducing unexpected interactions with higher-layer Congestion Control (CC) algorithms; hence, making their accurate evaluation in such environments an open and challenging task.

Moreover, today's Internet hosts an expanding oeuvre of CC algorithms, such as New Reno at Netflix [58], Copa at Facebook [7], and BBR [12] at Google.

However, the lack of widespread access to high-quality 5G prototype environments limits the ability of researchers to test their protocols' designs in natural 5G environments. Even when such access is available [16], it suffers from reproducibility and control issues. In the context of 5G networks, the research community lacks an end-to-end framework that enables effective testing of a variety of state-of-the-art protocols, in a repeatable and reproducible manner, due to multiple factors [40]. First, the simulation environment needs to model realistic fine-grained bandwidth, loss, and buffer variations. Second, many newly proposed CC solutions have tailored real-world implementations that lag in developing their corresponding simulation counterparts. Third, simulations and real-world evaluations may not always match due to protocol over-simplifications or experimental variations. This is due to the complexity of the lower-layer simulation models (especially PHY and MAC layers), whose behavior is modeled in order of milli—and micro—seconds, while much longer simulations are required to evaluate CC algorithms (i.e., order of tens of seconds).

Considering the challenges above, it is still necessary to understand the expected behavior of different CC algorithms over the potential scenarios brought by the 5G technology and analyze whether the current CC mechanisms can rise to the 5G challenge. This paper provides a detailed performance analysis of the ten most relevant CC algorithms over 5G scenarios. Such analysis includes single CC evaluation, and the interplay between different CC alternatives. To overcome the limitations of existing evaluation approaches, the analysis carried out in this paper was performed over a novel toolkit, whose design is also presented in the paper, called *Zeus*[1]. It supports multi-Gigabit traffic rate using an enhanced implementation of the Mahimahi [42] link emulator.

This work makes the following contributions:

- Throughput and delay comparisons of up to ten CC protocols over 5G channels, including real-life cellular 5G connections, mmWave simulated 5G channels, and WiGig (802.11ad) communications.

- Analysis of the shared buffer impact on CC algorithms including their individual and co-existing performance in the 5G environment.

- Adopting of the *harm* metric [63], to show the detriment caused to a CC algorithm performance by other co-existing flows over a shared channel.

- Design and implementation details of the *Zeus* toolkit, which has the following features:
    - It provides a framework for systematic evaluation of CC solutions so that they can be fairly compared over different scenarios, embracing arbitrary access technologies;
    - It embeds an enhanced Mahimahi implementation to enable CC protocol testing at multi-Gigabit data rates.

- Open-sourcing Zeus to the research community packed with several real measured 5G traces.

Based on our analysis of ten different congestion control protocols across various 5G network environments, we observe several interesting behavioral characteristics of these protocols. A few salient observations include: (i) BBR [12] is highly impacted when co-existing with Cubic, where its performance is highly dependent on the bottleneck buffer size. (ii) In the evaluated scenarios, Vivace [15] and Verus [66] exhibit high-performance variability over multiple runs under different 5G conditions; (ii) The throughput of delay conscious CC protocols, like Copa and Ledbat, is remarkably jeopardized in the process of holding delays at a specific threshold in 5G scenarios. (iv) Allegro [14] avoids Cubic's delay deficiencies and appears to provide higher throughput than competing solutions; however, it is still unable to achieve the full capacity of the 5G channels. (v) Online learning algorithms like Vivace and PCC-Proteus exhibit convergence issues due to the difficulties to identify a better operating point in highly variable 5G environments. (vi) None of the CC algorithms are suitable to co-exist without negatively harming each other's performance in the 5G environment.

## 2 Challenges in 5G networks

This section briefly outlines specific challenges to developing an experimental environment for a realistic evaluation of CC protocols in 5G and beyond-5G networks.

**High bandwidth variations and loss rates:**
5G networks can exhibit high bandwidth variations over short time scales, especially for high-frequency bands such as mmWave. Transmission rates reaching 8 Gbps [55] and 176 Gbps [19, 67] are observed within a 60 GHz frequency band in several scenarios. Moreover, mmWave is envisioned as a low-cost substitute for expensive fiber optic backhaul in cellular networks [32, 59], without compromising the multi-gigabit transmission rates [44]. mmWave signals suffer notable propagation losses when penetrating buildings or solid materials like walls. Besides the signal attenuation, the limited propagation distance can cause transmission outages or abrupt connectivity terminations when shifting from Line-of-Sight (LoS) to Non LoS (NLoS) conditions and vice versa [56] [25]. The emulation environment must accurately capture and support these bandwidth and loss variations over short time scales while allowing multi-Gigabit transmission capacity.

**Short-term link interruptions:**
Repeated LoS-NLoS state switching in 5G networks can trigger link disruptions over short time scales. In 4G and 5G protocol stacks, the Hybrid Automatic Repeat reQuest (HARQ) [54] configured at the MAC layer, and the acknowledged and unacknowledged modes (AM/UM) [34] configured at the Radio Link Control (RLC), use re-transmissions to compensate from packet losses. However, they cannot compensate for link interruptions triggered by sudden NLoS transitions that occur over time scales larger than a few round-trip time (RTT) periods. Such short-term link interruptions should be accurately captured in the emulation environment.

**Buffer sizes and bloats:**
Configuring appropriate buffer sizes in 5G networks is not easy. A typical approach to compensate for packet losses is over-provisioning network buffers. Buffer sizes above one bandwidth-delay product (BDP) in 3G/4G networks are considered to mitigate packet losses [31]. Such oversized buffers also create buffer bloats, leading to increased communication delays [64]. The emulation framework needs to support the ability to configure and control buffer lengths to capture different protocol variations under such environments.

---
[1] The toolkit will be open-sourced and released in a public GitHub repository once the work has been accepted to ensure the anonymity of the authors.

**Tracking available bandwidth:**
5G networks require protocols with fast rate adaptation for good performance. Many end-to-end congestion control protocols apply some form of bandwidth probing to derive the achievable data rate, such as in [5, 12, 13], and [47]. The proposed framework should support fine-grained emulation of different forms of bandwidth probing that protocols may adopt, and accurately capture variations in their performance in such network conditions.

**5G testing:**
Practical 5G cellular network experimentation in field trials is reserved exclusively for the network operators with considerable research and development (R&D) funds. Conducting field trials on a large scale is time-consuming and expensive. Today, many companies offer off-the-shelf testbed solutions [60], [6], [27], which enable accelerated R&D; however, they are experiment-specific and allow experiments mainly on the physical and MAC layers with no system-level R&D flexibility. This hinders their use because it prevents CC protocol developers from obtaining quick insights into their protocol performances. In this sense, the proposed CC evaluation environment provides some abstraction of the underlying technologies and protocols so that the main focus is on the CC performance evaluation.

## 3 The Performance Measurement Framework (ZEUS)

This section describes *Zeus*, the toolkit we designed and developed to conduct the performance analysis of CC protocols and algorithms over different 5G channels. Note that the analysis methodology described in this section is general and technology-agnostic, so that it can be extended to study the performance of CC algorithms over other types of channels.

The overall workflow is detailed in Figure 1. The framework permits sending traffic using up to 10 different CC solutions (kernel-based or application-based) with their actual implementation at high data rates and different scenarios. To emulate these scenarios and support such high data rates, the framework embeds a modified version of the Mahimahi [42] link emulator. This tool has been adopted by many works to measure the performance of CC solutions over various types of network conditions [4, 7, 15, 20, 65].

As for the channel traces, different sources can be used, as far as they follow the appropriate format, which is detailed in Section 3.2. Furthermore, the framework includes a set of traces obtained from three sources: a real 5G cellular network, a WiGig router connection, and mmWave traces generated with an ns-3 network simulator. *Zeus* embeds a tailored version of ns-3 [51] to generate simulated traces for reference mmWave scenarios, using the mmWave module [40] developed by New York University wireless group (NYU Wireless).

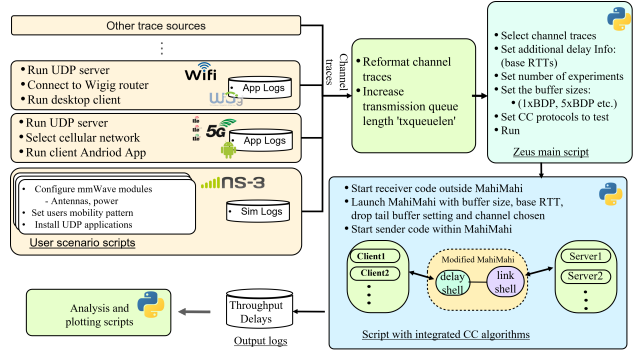

Figure 1: Architecture of the Zeus framework

In short, we send UDP bulk data over a given scenario and record the receptions to create the capacity traces. *Zeus's* main script (top right of Figure 1) chooses the traces that are used to model the channel dynamics and correspondingly configures the Mahimahi link. As shown in Figure 1, the link emulator can be further tailored with additional end-to-end delay and/or modifying the buffer size of the bottleneck link. Regardless of the chosen channel trace, Mahimahi is used by the framework to establish the bottleneck link between end-to-end connections. The framework sets up sender/receiver pairs using a CC solution and connects them through Mahimahi. The sender/receiver pairs then start sending traffic, and the framework generates result files, including packet capture (PCAP) files, logs of sending and receiving events, and evolution of buffer occupancy. Altogether, the output files permit us to compare the behavior of different CC solutions over the same circumstances in a systematic manner. Besides, *Zeus* also allows the analysis of bottleneck link-sharing among traffic flows, using different CC protocols. We have adopted these tools to leverage the best from real-life and emulation realms without the reproducible limitations of hardware-based studies.

In the following we detail the three most relevant aspects of the Zeus toolkit. First we describe methodology followed to generate traces from real 5G networks. The reader may refer to Appendix A for a more detailed explanation of the methodology developed to generate traces using the Wigig router and ns-3 simulator. Then, we detail the changes in Mahimahi to enable multi-Gigabit transmission, Finally, we compare the performance brought by the Zeus toolkit with that obtained over a real 5G network under similar circumstances.

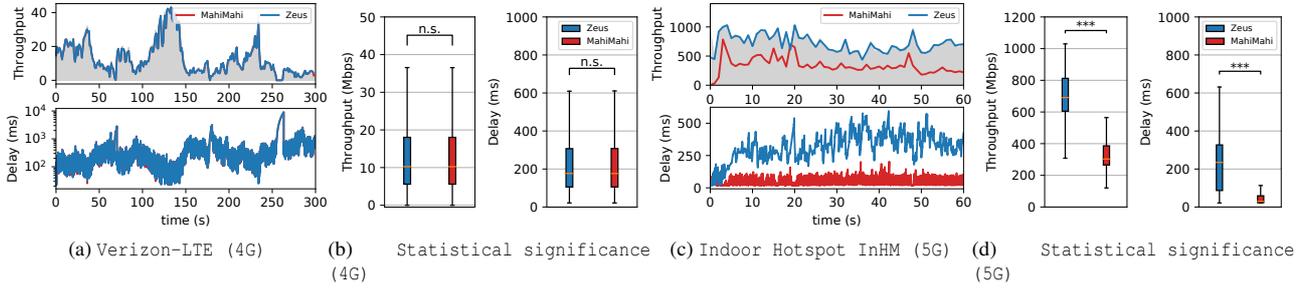

Figure 2: Benchmarking Zeus against Mahimahi on 4G and 5G channels

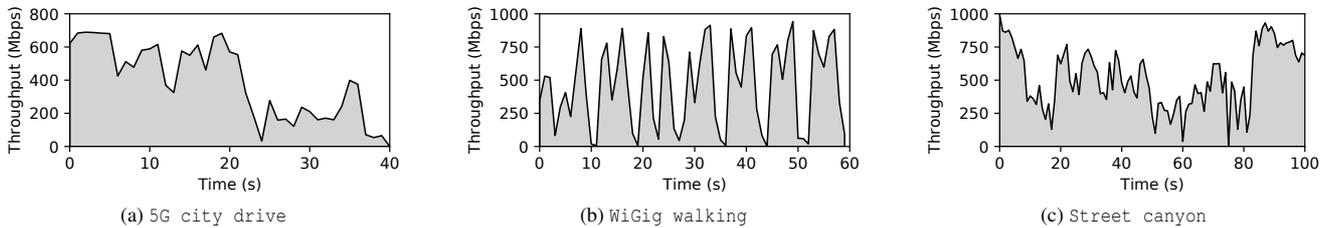

Figure 3: Evaluated 5G channel traces

### 3.1 Real 5G celullar network traces

For actual 5G network channel traces, *Zeus* offers a client and server-side implementation which can record real-time 5G channels in a cellular environment. However, the server must have a Gigabit upload capability upon a client connection. Usually, Internet Server Providers (ISPs) advertise Internet plans, with download speeds reaching Gigabits per second (Gb/sec). However, the upload data rate is often capped at a threshold for all online activities. This threshold varies with regions and ISPs available in the area. Connecting the server to the ISP's backbone network via a fiber link with Gb/sec upload and download capacity is recommended. The client consists of an Android application that can communicate with a server using the server's IP and port number. Upon connection, the Linux-based server begins sending UDP packets with a maximum transmission unit (MTU) size of 1500 bytes at a constant data rate ($\geq$ Gb/sec) to the IP address of the Android mobile client. The Android application serves as a sink and logs the inter-arrival times of the received UDP packets. These logs are then utilized to create a trace file and converted into a proper channel trace format for the modified Mahimahi emulator explained below.

We recorded real 5G network channel traces by considering five different scenarios. Using Zeus's Andriod app, we logged packet arrival timestamps in an indoor and outdoor car park, slow driving in the city, walking along the beach, and sitting in a public park. These real-world channel traces using 5G cellular network are referred to as: Indoor car park, Outdoor car park, City drive, Beachfront walking, and Public park, respectively, in this paper. On average, each trace recorded for 60 seconds consumed 16 GigaBytes (GB) of cellular data.

### 3.2 Modified Mahimahi Emulator

Among the different functionalities of Mahimahi, we mainly use its capability to replay channel traces (i.e., mm-link tool). However, the original Mahimahi implementation has certain limitations that affect the emulation of multi-Gigabit capacity channels, such as those based on mmWave. The original implementation has certain bottlenecks that result in a limit of around 400 Mbps on the throughput that the network can handle efficiently. The developers acknowledge that Mahimahi is not tested beyond 100 Mbps. Any trace exceeding the threshold of $\approx$ 400 Mbps causes Mahimahi to drop packets randomly, thus capping the channel throughput and utilization.

To address this issue, we made several modifications, some after discussions with the original developers. First, we reduced the number of read and write system calls. Mahimahi's mm-link program simulates a channel based on a trace file, which contains one number per line. Each number indicates the millisecond from the channel's start time to accommodate a packet. Thus, the trace file shows the channel's capacity regarding the number of packets accommodated every millisecond. The mm-link program simulates the channel by holding each incoming packet and releasing them at the specified millisecond mark in the trace file. The original Mahimahi makes a call to read and write for each packet. We speed this up separately for the `read` and `write` calls. For the `write`, all the packets are read in every millisecond from the trace file and make a single call to `write` when writing to the internal socket. For the `read`, we make the socket non-blocking and read in as many available packets in a single instance before buffering them. Another change was increasing the transmission queue length (`txqueuelen`) of the virtual network interface from the original value of 1000 to 500000 `Bytes`. This latter change aims to avoid overflow of the internal interface queue with multi-Gigabit traffic rate. Together, these set of changes significantly increased Mahimahi's throughput, allowing experiments over multi-Gigabit capacity traces.

Figure 2 shows a performance comparison of the original Mahimahi implementation and our modified version in *Zeus*. The results are obtained by having a single Cubic flow with bulk full-buffer transmission during the whole experiment duration. Figures 2a and 2b show the results obtained with a 4G Verizon-LTE trace provided by [64], whose maximum capacity is around `40Mbps`. Figures 2c and 2d show the performance obtained over a synthetic 5G channel generated in ns-3 using Indoor Hotspot Model (`InHM`), with capacity reaching `1Gbps`. Both channel capacities are shown in Figures 2a and 2c with the shaded background. It is evident that for the 4G Verizon-LTE channel, both Mahimahi and Zeus offer the same performance, with no statistical significance observed in terms of throughput and delay between the two, completely saturating the channel capacity (Please refer to §4 (iv) for statistical significance). However, for the `InHM` 5G channel, the original Mahimahi version fails to handle the number of sending events leading to the under-utilization of the channel capacity, and eventual packet losses due to buffer overflow. In contrast, *Zeus* manages to support all the generated traffic, enforcing the correct emulation of the 5G channel, thus mimicking the correct behavior of the CC protocol in use. This is done by significantly lowering the required inter-process events for such high-bandwidth 5G traces. A considerable statistical difference for both the throughput and delay is observed in the performance comparison as illustrated by Figure 2d.

## 3.3 Zeus' operation validation

We validate the correct operation of *Zeus* by comparing its performance with that obtained in a real environment. In particular, in Figure 16 we represent the statistical distribution of the delay and throughput with whisker plots when using *BBR* as CC solution. The delay is measured as the time elapsed since a packet leaves the sender until an Acknowledgement (Ack) is received back at the sender, confirming the packet delivery. In contrast, the throughput is periodically measured every second, taking the number of bits received within the last second. In Figure 16 we show the results obtained from 3 5G cellular connections, each lasting 60 seconds, using

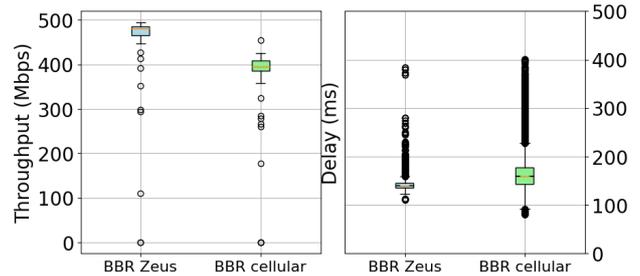

Figure 4: Throughput and delay comparison of *BBR* in a real 5G cellular connection Vs. *Zeus*

*BBR* during the connections, and results obtained by *Zeus* for *BBR* over similar channel traces recorded at the exact location where the 3 5G cellular connections are created.

In Figure 16 we can see that the results obtained with *Zeus* are similar to those obtained from the actual 5G connections, both in terms of average value and sparsity. In particular, the throughput obtained in both cases, *Zeus* and cellular, reach comparable average values and tight distribution. In the case of the delay, the average values are again alike, while the sparsity observed for the *BBR* cellular connection is slightly larger. Although the results are not identical, they serve to ensure that the CC protocols experience similar and comparable conditions with *Zeus* and an actual cellular connection. Obtaining identical results would not be possible, since the wireless channel realizations are different for the 5G connection running *BBR* and those using UDP to generate the traces used by *Zeus*.

## 4 Evaluation

Our evaluation aims to explore the performance of state-of-the-art CC protocols from the following perspectives:

**(i) Diverse 5G networks**:
Initial 5G roll-outs heed an NSA flavor of 5G built on existing 4G-LTE Evolved Packet Core (EPC). It is an upgrade to quickly introduce partial 5G services while maximizing the reuse of existing 4G networks, helping Mobile Network Operators (MNOs) from suffering economic impediments. Moreover, the adoption of mmWave into 5G systems is an essential component currently missing in the 5G puzzle. Although 3GPP has already finalized 5G NR Release 16, the actual deployment of mmWave technology making its first public debut in 5G is still unknown. Therefore, we include three types of networks in our evaluation: a) an existing NSA 5G cellular network, b) a physical 5G WiGig router for a 5G mmWave network, and c) an ns-3 mmWave network. The *Zeus* toolkit is packed with various channel traces that we recorded over diverse 5G scenarios. Due to space limitations, we selected three distinct channel traces obtained from a real 5G cellular network (5G city drive), a physical 5G WiGig router (WiGig walking), and a simulated mmWave ns-3 scenario (street canyon) for an in-depth performance evaluation. Figure 3 illustrates the selected 5G traces, with their varying capacities over time.

**(ii) End-to-end throughput and delay**:
It is sensible to question the compatibility of existing congestion control mechanisms with the characteristics of 5G networks. CC protocol performances may encounter attrition factors earlier highlighted in §2. Therefore, we measure the end-to-end performances of CC protocols to identify their anomalies in the 5G domain.

**(iii) Harm analysis of CC protocols**:
It has become a norm in the existing literature to establish the fairness of new CC algorithms towards existing legacy CC solutions while sharing a bottleneck bandwidth. Several papers studying CC techniques exploit Jain's fairness index to ascertain this aspect. However, unlike fairness or friendliness to legacy TCP, in our analysis, we adopt a more practical harm-based approach (defined in [63]), which aims at identifying whether a new CC algorithm causes more harm to the performance of an existing legacy CC protocol when sharing a 5G bottleneck than the legacy protocol causes to itself. This rationale is to ensure that new techniques do not cause more harm than existing solutions. Taking Cubic as our default TCP flavor[2], we measure the harm caused to Cubic flows by other competing flows of SoTA CC protocols. To identify if a new protocol is eligible to co-exist with Cubic in the wild, it must not damage Cubic flows beyond the damage Cubic causes to itself. In simple words, the maximum harm allowed for a new CC protocol to be deployed alongside Cubic should not surpass Cubic's self-harm.

**(iv) Statistical Significance:**
Statistical significance establishes how confident we are that the difference or relationship between two random variables does not exist by chance. When a result is identified as statistically significant, there is an absolute difference or relationship between two variables, and it is unlikely that it is a one-off occurrence. In our case, we are interested in measuring how different CC solutions are. More specifically, two variables are statistically significant when their p-value falls below a certain threshold, called the *significance level*. We represent the level of significance with stars. If the p-value is less than 0.05, the level of significance is flagged with one star (*). If a p-value is less than 0.01, it is flagged with two stars (**). If a p-value is less than 0.001, it is flagged with three stars (***). If the p-value is larger than 0.05, then there is no statistical significance between the two variables, and it is flagged with (n.s). Our evaluation tends to identify the statistical significance in terms of throughput and delay of CC protocols compared to Cubic.

---

[2]We choose TCP Cubic as our baseline reference protocol since it is the default TCP flavor in many of today's platforms such as Linux [64], or Android OS [3].

**Setup:**
All experiments are set with a propagation delay of `10ms`, resulting in a `20ms` minimum RTT. The experiments are conducted on a customized server with an Intel Xeon Bronze 3204 CPU @ 1.90GHz × 12, a 15 GiB memory, and an Ubuntu 20.04.3 LTS Operating system. Note that for Gbps data transfer, we fine-tuned the server. A complete instruction guide on server tuning is packed in the Zeus[1] toolkit. The experiments are performed with a bftpd server and an FTP client, requesting a large file from the server using a given CC algorithm. For each CC solution-channel tuple, we ran 20 independent experiments. To avoid unfair performance logs for all CC solutions, we used a file size of 20 GB to ensure traffic throughout the experiment duration (`180sec`)

## 4.1 End-end throughput and delay analysis

**Legacy CC protocols performance (Cubic & Reno)**
Figures 5a, 6a, and 7a present scatter plots (throughput Vs. delay) of ten SotA CC protocols over a real 5G cellular trace, a physical WiGig router trace, and the ns-3 street canyon trace, respectively. Each marker point represents the average throughput (y-axis) and delay (x-axis) performance for a CC solution over a single run. For each CC protocol, we conducted 20 separate runs per channel trace with an infinite buffer setting in *Zeus* to ascertain the stability of protocols' performance and how efficiently they can utilize the actual 5G network capacity. We observed that, across all three 5G traces, Cubic and Reno managed to almost saturate the channel capacity (average channel capacity is shown in the plots with a red dotted horizontal line). However, in comparison, both Cubic and Reno account for the highest delays among almost all the other CC protocols. We observe no statistical significance (n.s) in terms of throughput between Cubic and Reno for the WiGig walking and street canyon traces (Figures 6b and 7b, respectively). For the 5G city drive trace, the throughput comparison is flagged with two stars, highlighting a medium statistical significance in the results. In contrast, there is no statistical significance in delay for the 5G city drive trace (Figure 5c). For delay comparison of the WiGig walking and the street canyon traces, we observed a statistical significance in the results of the two protocols, highlighted by the two and three stars, respectively (Figures 6c and 7c).

**Google's BBR performance**
BBR shows very stable performance across all traces, almost fully reaching the channel capacity. Particularly, its throughput remains slightly lower than the average channel throughput, with 94% utilization over the 5G city drive trace and 92% utilization for both the WiGig walking and street canyon traces. Despite BBR's small loss in throughput utilization, it makes up for it in the delay performance, yielding a remarkable reduction in the delays' performance by one-half, one-fifth, and one-third times, compared to both Cubic and

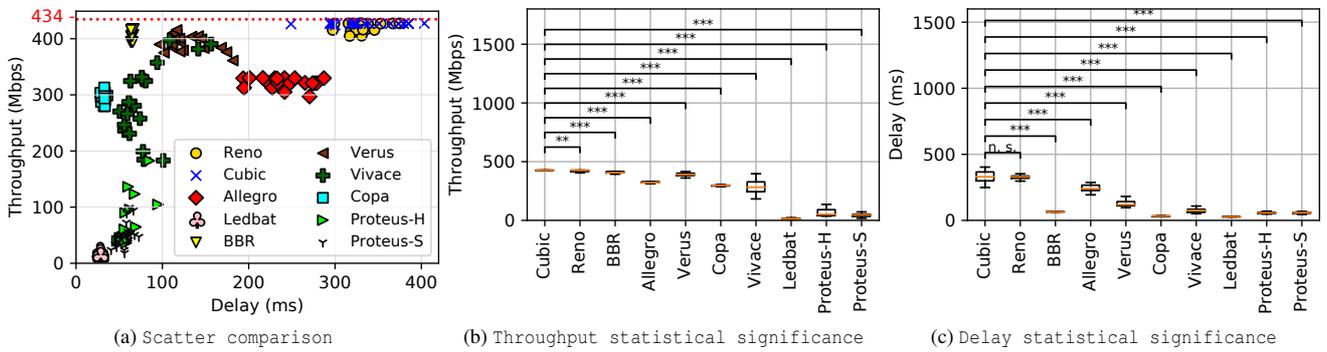

Figure 5: 5G city drive CC protocols comparison

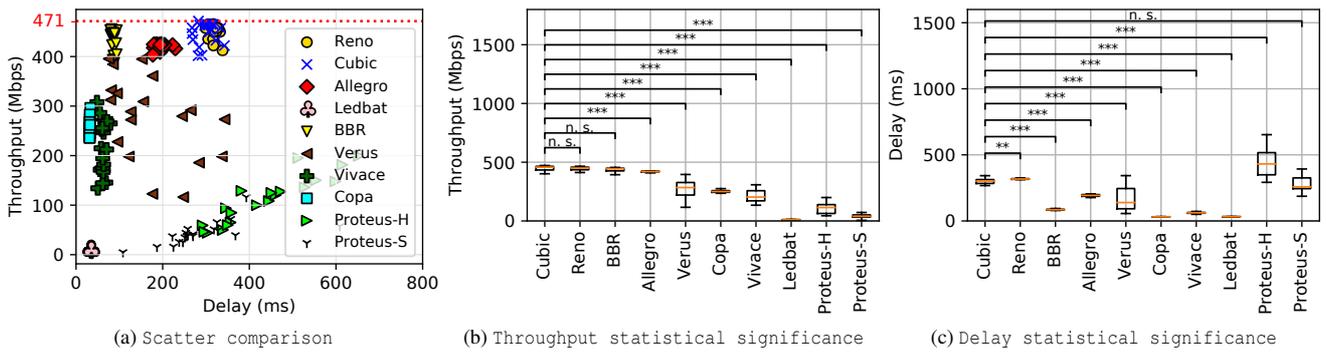

Figure 6: WiGig walking CC protocols comparison

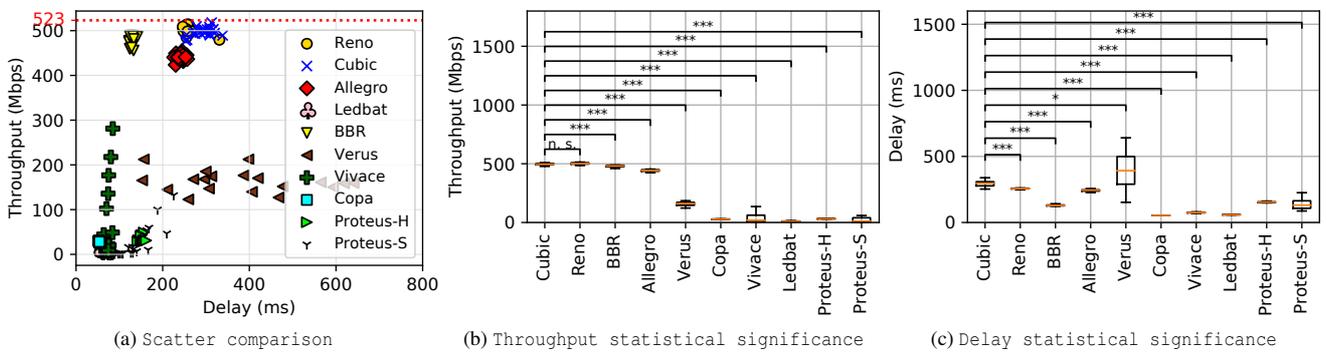

Figure 7: Street canyon CC protocols comparison

Reno for the WiGig walking, 5G city drive, and street canyon traces, respectively.

Looking at the statistical significance, it can be seen that for both the 5G city drive and the street canyon traces, there is a statistical significance (***) in the results compared to Cubic (Figures 5b, and 7b). However, for the WiGig walking trace, no statistical significance of the throughput is observed between BBR and Cubic (Figure 6b). On the other hand, BBR's delays are of statistical significance (***) when compared to Cubic, across all the three traces (Figures 5c, 6c, and 7c), which highlights their different expected behavior in terms of delay in all scenarios.

**Facebook's Copa performance**
Copa maintains a stable behavior for all traces, although its performance is degraded in terms of throughput compared with the aforementioned alternatives. Copa's throughput varies across the different channels despite keeping the delays at a minimum. As seen in Figure 7a in the street canyon trace, Copa manages to explore only 5% of the available channel capacity, whereas it achieves a 68% and 54% channel utilization for the 5G city drive and WiGig walking trace, respectively. Aligned to that, Copa's performance is statistically significant compared to Cubic, in terms of both delay and throughput.

**The PCC Family performance**
Allegro achieves lower delays than Reno and Cubic but higher delays than BBR. Apart from the 5G city drive trace, Allegro's throughput performance is higher than all CC protocols but BBR, Cubic, and Reno. In all cases, Allegro maintains its delay within the `200-300ms` range. Note that the x-axis of Figure 5 is from `0-400ms` whereas, for the other two traces, it is set from `0-800ms` to capture the whole range of delays. As for Vivace, it shows steady results in terms of delay, although it exhibits a highly variable throughput in independent experiments over the same channel trace. In Figure 5, we observe that Vivace manages to achieve a high throughput, whereas it is not able to explore the channel capacity in the other scenarios.

Following a scavenger strategy, Proteus-S is not able to adapt to variability of the analyzed channels, reaching only 3% of the available capacity. Similarly, Proteus-H, a hybrid CC that switches between Proteus-S and Vivace, exhibits similar behavior to vivace and Proteus-S. However, in terms of throughput and delay, its behavior is volatile. Allegro, Vivace, and Proteus-H have a high statistical significance for delay compared to Cubic across all traces, given their substantially lower delays. However, this comes at the expense of throughput, where the Cubic results are of statistical significance. As for Proteus-S, there is no statistical significance in the delay results compared to Cubic in the WiGig trace, although with a substantial loss in throughput.

**Others**
Observing the Verus results, it exhibits scattered throughput and delay values across all traces, making it unpredictable protocol in 5G environments (specifically in the WiGig and street canyon traces). Similarly, Ledbat, which adopts a scavenger strategy like Proteus-S and fails to saturate the link capacity, achieves low throughput in all traces.

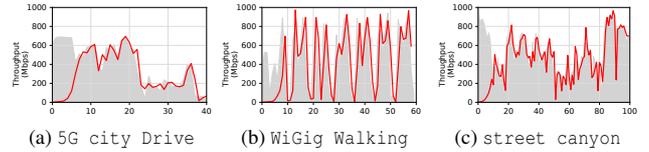

(a) 5G city Drive  (b) WiGig Walking  (c) street canyon

Figure 8: Allegro's instantaneous throughput across the 5G traces.

## 4.2 Why are the protocols behaving this way?

Both Cubic and Reno are known for building up large sending windows, resulting in unnecessary delays due to over buffering packets within the networks. Such behavior often leads to a well-known phenomenon called "bufferbloat" in cellular networks and has been well investigated in the context of 3G/4G cellular networks [31]. We observe the same behavior with an infinite buffer in the 5G network. Both Cubic and Reno rev the sending rate saturating the available channel capacity resulting in higher packet delays. Moreover, BBR's slight gap in full utilization of the channel capacity is due to its operation in a series of states, i.e., the startup phase, a PROBE_BW phase every 8 RTTs to estimate bandwidth, and a PROBE_RTT phase every `10 sec` to re-estimate the minimum RTT. The Probe RTT phase is when BBR briefly reduces its packets in-flight to just four packets, draining any queue built up, and it appears to be the cause of the slight under-utilization of the available channel. Copa also relies on queuing delays for congestion control and strives to minimize the packet queuing delays by periodically draining the buffers. However, Copa's effort to maintain a minimum queuing delay results in a significant throughput degradation in 5G networks.

Allegro's primary weakness is its fixed-step increment or decrement in sending rate. We observe these steps may be too small in a 5G network setting. As observed in the instantaneous throughput plots given in Figures 8a, 8b, and 8c, Allegro seeks to saturate the link capacity in steps at the startup phase. However, when the sending rate becomes significantly higher than the channel capacity, Allegro drops its sending rate in fixed increments until it reaches the bandwidth threshold again. This fixed-step sending rate limits Allegro's ability to achieve higher channel utilization and comes at the cost of increased delays.

As admitted by its authors [15], Vivace does not perform well in a highly dynamic network such as cellular networks. Vivace is a learning-based CC algorithm using online greedy optimization methods to control its sending rate via a utility function. However, optimization methods are easy to trap in a local optima [36], which is why we see a different throughput performance with repeated experiments. Moreover, Vivace desires to approximate the lowest achievable RTT (RTT$_{min}$) to decide its sending rate, which explains the steady delays. However, in 5G networks, operating at RTT$_{min}$ is not necessarily optimal due to drastic changes in available bandwidth even within a single RTT, making Vivace not an appropriate protocol for bandwidth-sensitive 5G applications.

Proteus-S and Proteus-H are built atop an online learning CC framework [14]. Proteus-S, with its scavenger utility, controls sending rate and leverages delay variation as a sensitive early indication of flow competition. Despite having a strategy to compensate for misleading delay variation in a non-congested channel (which is highly likely in 5G channels), it still exhibits low performance. We think this is because of the uncertainty of the learning-based components and the difficulties to identify the change of the equilibrium points or better operating point. On the other hand, Proteus-H controls its sending rate by appropriately choosing between the utility function of Proteus-S and a modified utility function of Vivace (with negative RTT gradients ignored); it inherits the drawbacks of both Proteus-S and Vivace.

Ledbat tries to keep delays at a predefined target value (default=25ms), and stays at the value to form a steady-state and never add more than a target delay in queuing. Ledbat's congestion signal is RTT exceeding a target threshold. This explains why Ledbat achieves a low throughput and delay across all channel traces. Whereas Verus achieves variable performance in the 5G scenarios. Even on the same channel, repeated experiment results vary significantly. We attribute this to Verus's delay profile curve mechanism, which is its main mechanism behind the congestion window decisions. From the results, it seems that Verus learns a different delay profile curve in each experiment to alter its behavior. However, it seems that this mechanism is not fast enough to catch up with the highly variable 5G channels and causes the performance to vary significantly from one run to the other.

### 4.3 Harm Analysis

A common requirement for any new CC protocol is its fairness in sharing the bottleneck bandwidth with existing TCP (e.g., Cubic). However, this goal is too idealistic to execute in practice. It is believed that being unfair to Cubic is acceptable because Cubic is not even fair to itself [7]. In addition, Jain's Fairness Index [30] used to measure fairness assigns an equal score in both cases, i.e., whether a new CC protocol utilizes a larger share of the bottleneck bandwidth than the legacy Cubic or vise-versa. Therefore, inspired by [63], we follow a harm-based approach in this analysis to identify the damage/harm the CC algorithms do to Cubic when they co-exist with Cubic over the 5G channels. We chose BBR, Reno, and Copa for our analysis because of their deployment in today's most used and popular services: Google, Netflix, and Facebook. We also considered selecting Allegro and Verus based on our end-to-end throughput and delays analysis. Following the harm definition in [63]), we define x= solo performance; and y = performance after introduction of a competitor connection. Then for metrics where 'more is better' (e.g., throughput) the harm = $\frac{x-y}{x} * 100$. On the other hand, for metrics where 'less is better' (e.g., delay) harm = $\frac{y-x}{x} * 100$.

We conducted 20 experiments with an "infinite buffer" setting for each trace with two Cubic flows co-existing on a 5G channel to calculate Cubic's throughput and delay self-harm as the reference threshold (i.e., a baseline for comparison). This threshold shall provide a firm definition for when a CC protocol can be deployed alongside Cubic (i.e. it does not produce more harm than Cubic itself) and when it is not. We represent this threshold in the horizontal dark green areas shown in Figures 9a, 9b, 10a, 10b, 11a, and 11b. The green shaded area in the plot represents the safe area reflecting the acceptability of the protocol alongside Cubic when sharing a bottleneck link. In contrast, the red part reflects the zone where the protocol is not suitable for deployment alongside Cubic. We observe that Cubic causes 50% throughput-harm to itself, reflecting an equal sharing bandwidth behavior when competing with another Cubic flow. Note that negative delay harm indicates a reduction in Cubic delays, whereas a positive delay-harm shows an increase in delay.

In all traces, we observe that BBR, Allegro and Copa falls in the green zone not doing much throughput or delay harm to Cubic. However, Reno is the only CC that always falls into the red area and harms Cubic in terms of both throughput and delay compared to other CC protocols. Reno does more than 80% throughput-harm with 60% delay-harm to Cubic in the 5G city drive trace, above 60% throughput-harm and nearly 90% delay-harm in the WiGig walking trace, and 55% throughput-harm and nearly 100% delay-harm in the street canyon trace. This could also be why Netflix adheres to Reno due to its aggressive behavior in dominating other CC protocols. However, Google announced that Netflix is currently experimenting with BBR [13]. Although BBR is expected to make a major departure from traditional congestion-window-based CC; however, it does not harm Cubic's performance and allows Cubic to take most of the channel capacity across all traces. In Figures 9a, 10a shows that BBR does not have a strong impact on Cubic's throughput. In the case of the mmWave NS-3 channel, Figure 11a shows a slight impact of BBR over Cubic, but again below Cubic's self-harm threshold. Regarding delay, Figures 9b, 10b and 11b, BBR's impact is slightly higher, but in all cases it is below the Cubic's self-inflicted harm. We further analyze the interaction between Cubic and BBR in Figures 9c, 10c and 11c, where we observe that Cubic dominates the channel capacity, unfairly killing BBR flows in an "infinite buffer" setting. This leads us to investigate if the same behavior is valid with various buffer

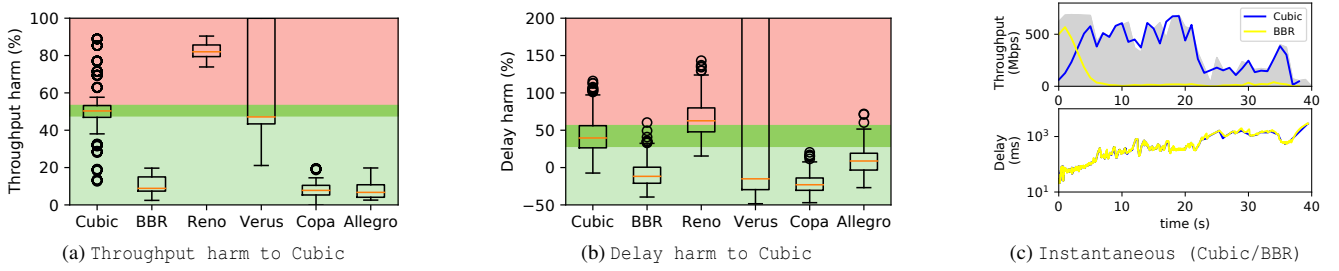

Figure 9: Comparison of CC protocols' performance over real 5G city driving channel trace with infinite buffer.

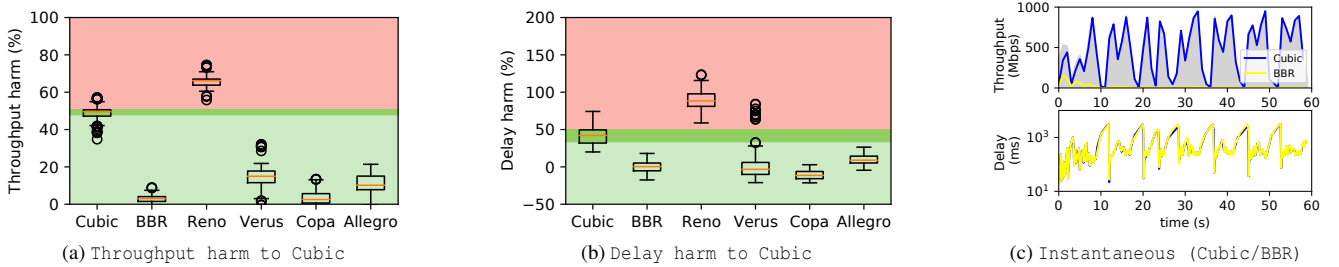

Figure 10: Comparison of CC protocols' performance over WiGig walking trace with infinite buffer.

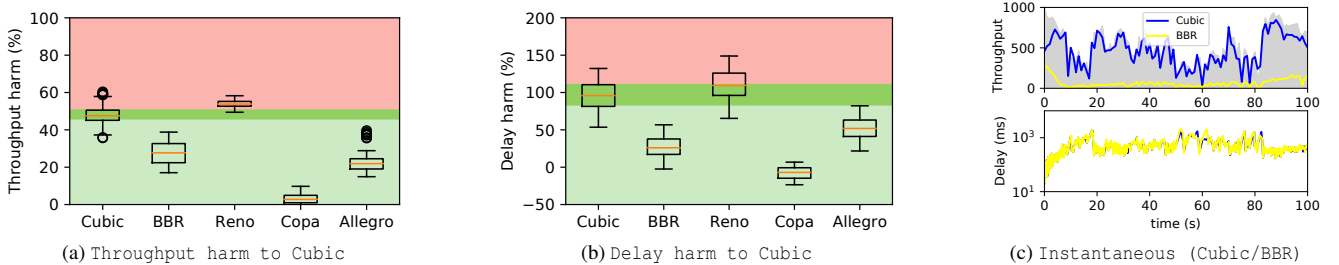

Figure 11: Comparison of CC protocols' performance over 5G street canyon trace with infinite buffer.

sizes.

### 4.4 Impact of buffer sizes

To understand BBR's behavior when sharing a bottleneck link with Cubic and assess the impact of the buffer size on its performance, we tested it with different buffer sizes, varying from 1 to 15 BDPs. Figures 12a and 12d show the throughput and delay inter-change between BBR and Cubic according to the bottleneck buffer size in the 5G cellular city drive scenario. The figures show that at approximately `4.5xBDP` both protocols reach a fair share of the capacity. On the other hand, Cubic claims more capacity upon increasing the buffer size, almost to the point of full dominance around `7xBDP` and above. As for values falling below the `4.5xBDP` size, BBR dominates the performance leaving almost no share for Cubic to explore.

Similarly, for the street canyon trace results shown in Figures 12c and 12f, both BBR and Cubic reach a fair share at around 11.5xBDP before Cubic dominating the channel capacity with growing buffer sizes. We find two logical explanations for such behavior. First, BBR restricts the packets in-flight at a maximum of `2xBDP` (the extra BDP deals with delayed/aggregated ACKs). As a result, the extra BDP of data in shallow buffers causes huge packet re-transmissions due to losses. BBR neglects loss as a congestion signal and maintains high re-transmissions over time, worsening things. On the other hand, Cubic adjusts its `CWND` upon detecting a loss. Therefore, BBR causes more packet transmission/re-transmissions than Cubic. This also implies that BBR delivers high throughput in shallow buffers but at the expense of high packet re-transmissions. Second, BBR finds the maximum `target_CWND` as `CWND_gain x BtlBw x RTprop` and increases the congestion window each time an ACK is received until the window reaches the target_CWND. Where BtlBw and RTprop are the estimated bottleneck bandwidth and round-trip propagation delay, respectively. Every 10 secs, BBR probes for RTprop by reducing its in-flight packets to just four packets to drain the queue. When BBR has an accurate estimate for the BtlBw and RTprop, it caps its in-flight packets at 2xBDP. This means that BBR allows just 1 BDP of packets to queue at the bottleneck buffer for 8 RTTs. Meanwhile, Cubic expands its CWND to fill the buffer before encountering loss. Since a flow's throughput is proportional to the buffer share, Cubic gets more packets queued. BBR observes a lower throughput and thus further decreases its CWND. This creates a positive feedback loop allowing Cubic to continue its CWND increase in response to BBR's decrease in CWND.

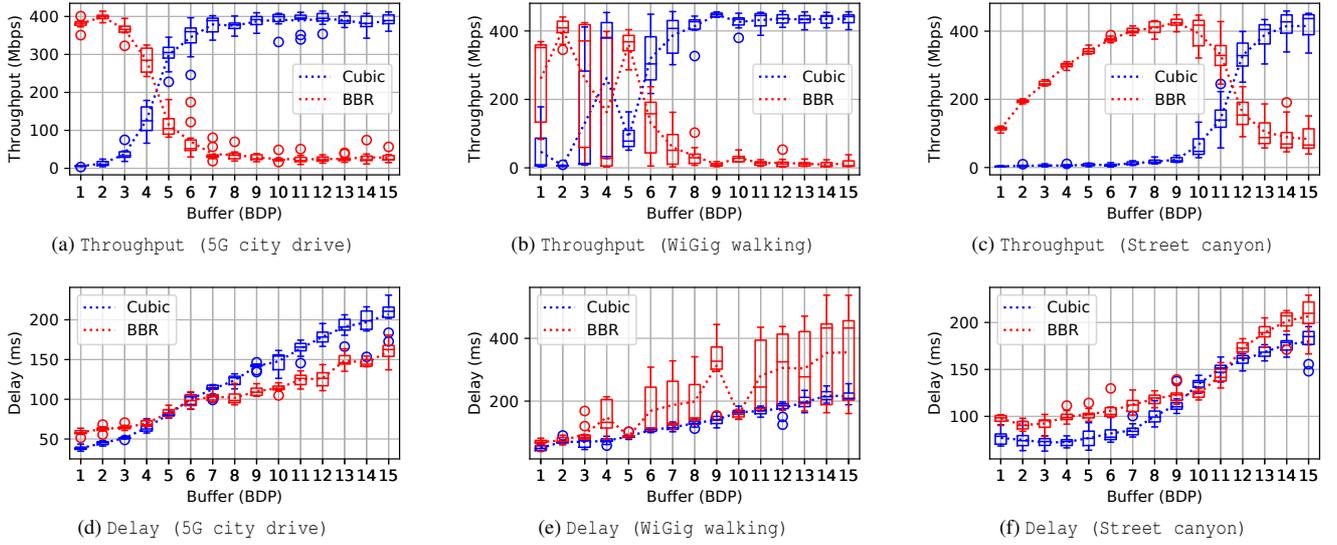

Figure 12: Effect of buffer size on Cubic and BBR sharing a bottleneck across the 5G channel traces.

Figures 12b and 12e however, show a different behavior over the WiGig walking trace. We know that Cubic being a loss-based CC, tends to saturate the bottleneck buffer up to exhaustion, regardless of its size, whereas BBR restricts the in-flight data to 2xBDP. This means that the larger the bottleneck buffer, the bigger share for Cubic. However, since Cubic builds long-lasting standing buffer queues, a competing BBR flow may not acquire the true $RTT_{min}$ during its PROBE_RTT phase. During the PROBE_RTT phase, BBR significantly reduces its in-flight packets to drain the queue; however, Cubic does not. Thus the bottleneck buffer is not drained completely, which makes BBR assume a higher $RTT_{min}$ and thus, it increases its in-flight cap to a larger value. Occasionally, BBR gets a nearly fair share of the bottleneck bandwidth. At other times, most of the bottleneck bandwidth is occupied by the Cubic flow.

This analysis forces us to presume that BBR might not always be an appropriate choice for 5G. For instance, if the content delivery is sensitive to packet loss and the buffers are shallow, one must carefully examine the trade-offs between throughput and Quality of Experience (QoE) before deciding on BBR as the main CC option. Additionally, from the results presented above, there is no clear pattern on what exact buffer size should be maintained for BBR to be harmless to Cubic since we have a different cut-off size in each trace. This makes deploying BBR in practice over 5G cellular channels very difficult. Consequently, we extend our analysis to determine whether other CC protocols can co-exist with the Cubic and identify their impact on Cubic's performance under different buffer settings.

### 4.5 Harm analysis with different buffer sizes

Figures 13, 14, and 15 show the harm analysis of the CC protocols chosen earlier with respect to Cubic in a shallow, medium and large buffer setting i.e., 2xBDP, 5xBDP, and 10xBDP, respectively over the 5G traces[3]. We observe that in a shallow buffer 2xBDP, BBR, Verus, and Allegro fall in the red zone harming Cubic in terms of throughput. We have noticed how BBR and Cubic co-exist in different buffer settings in Figure 12. A similar expected behavior in throughput and delay harm is observed here. In a shallow buffer of 2xBDP and 5G city drive trace, Reno equally shares the bandwidth with Cubic with no harm done to Cubic's throughput and delay. However, with 5xBDP and 10xBDP buffers Reno harm to Cubic increases to 65% and 75% in terms of throughput while Cubic delays

---
[3]We excluded assessing the Verus harm on Cubic from the street canyon trace due to Verus's highly unstable performance. We noticed that Verus, in most cases, failed to exit its start-up phase.

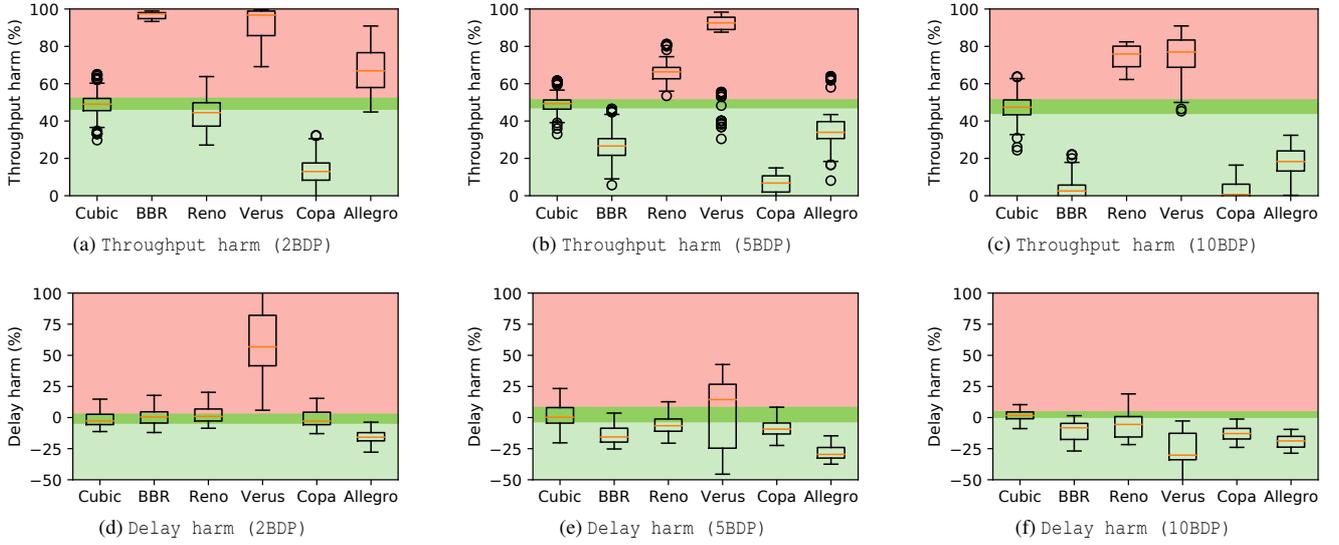

Figure 13: Effect of buffer size on the CC protocols' harm for the 5G city driving channel trace.

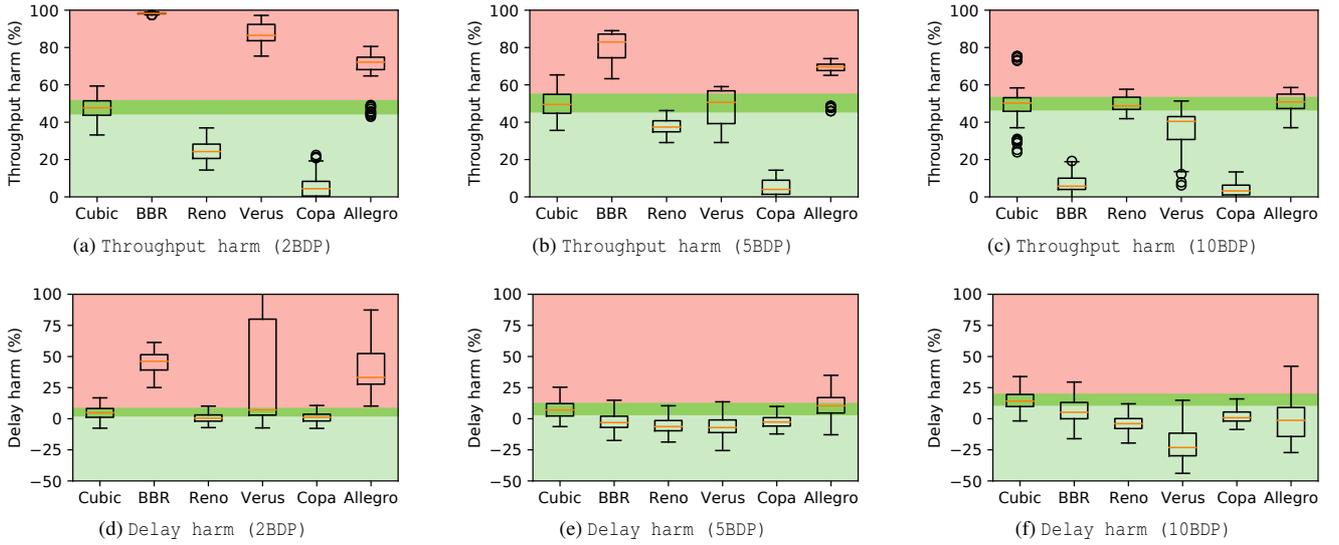

Figure 14: Effect of buffer size on the CC protocols' harm for the WiGig walking channel trace.

are reduced by 10% based on the observed (-ve 10%) delay-harm. Verus, with its uncertain behavior, harms Cubic above 95% in a shallow buffer and 80% in a `5xBDP` setting in terms of throughput. Copa remains in the green zone for all traces, and buffer sizes regarding throughput and delays harm due to its earlier low throughput performance. Allegro appears to be aggressive to Cubic in shallow buffers with nearly 75% throughput harm to Cubic across the traces. In medium-size buffers (`5XBDP`), Allegro's throughput harm to Cubic is 35%, 75%, and 80% for 5G city drive, WiGig walking, and street canyon trace, respectively. In contrast, the delay-harms remain in the permissible green zone. For `10xBDP` buffers, Allegro's throughput harm remains in the green area for 5G city drive and WiGig walking traces, whereas a 70% throughput harm in the street canyon trace. Allegro's delay-harms, however, remain in the green zone for `10xBDP`.

## 5 Related work

**Network Simulators and Emulators:** There is a wide variety of literature available on different simulation and emulation frameworks for cellular networks. Research on congestion control has long used network simulators like Network Simulator version-2, version-3 (NS-2, NS-3) [28, 51] and OMNeT++ [62], which can be used to virtually model any network. Moreover, real-time emulators like Dummynet [52], NetEm [26], Mininet [24], and MahiMahi [42] are broadly utilized for congestion control research. These run in realtime, limiting the scale of experiments to whatever the underlying hardware provides, while they exhibit adequate parameters and mechanisms to recreate complex network responses.However, setting up the emulator with realistic parameters to match the settings used in commercial networks is an open problem. The Pantheon project [65] is presumably the one that presents more similarities with *Zeus*. However, this project focuses more on live-testing and is less flexible for customized scenarios, specifically for 5G mmWave, hindering repeatability. Although Pantheon also provides MahiMahi as a tool for emulation, it does not support capacities above a specific limit ($\approx$ 500 Mbps). Moreover, the authors announced to no longer maintain Pantheon.

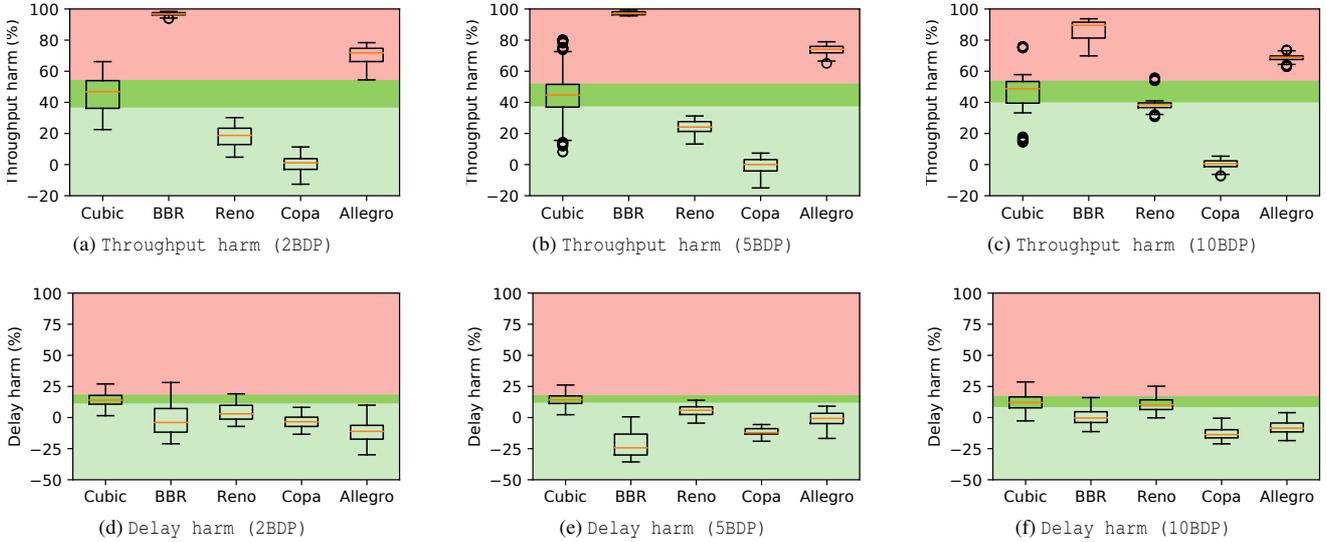

Figure 15: Effect of buffer size on the CC protocols' harm for the street canyon channel trace.

**Congestion Control:** Many CC protocol designs are based on assumptions about cellular bottlenecks and cross-traffic, and target specific scenarios that often restrict their deployment. Traditional TCP variants (e.g. [10, 22, 35]) fail to deliver consistent high performances [7, 12, 66] in wireless networks because they consider packet loss as a sign of congestion, which may not be accurate in wireless environments. Numerous CC protocols are proposed to address TCP's issues, which can be categorized as follows. The performance-oriented congestion control (PCC) family has contributed several CC protocols, leveraging a sending rate control logic based on an empirical utility function, such as PCC-Allegro [14] and PCC-Vivace [15]. One of the most recent proposals is PCC-Proteus [39], which can act as either a primary protocol (Proteus-P or PCC Vivace), as a scavenger (Proteus-S) or switch between both (Proteus-Hybrid) using a dedicated scavenger utility function. Recently, Google presented their BBR algorithm [12, 13] showing good performance over wireless channels with low latency. BBR uses an estimate of the available bottleneck link bandwidth and RTT to govern its pacing rate. Copa [7] leverages the observed minimum RTT to achieve a target rate that optimizes a natural throughput function and delay under a Markov packet arrival model. More on CC protocols and their performance analysis can be found in recent surveys such as [23] and [50].

**5G Testbeds:** The need for 5G testing frameworks has motivated academia, governments, and industry to collaborate [41] for the development and deployment of testbeds, each with its unique purpose. Some prominent examples of these testbeds deployed in Europe are 5TONIC [2], 5GIC [33], FOKUS [17], Aalto 5G network [1], and NITOS [43] and in the United States some recently deployed testbeds include TurboRAN [37], PhantomNet [8], COSMOS [49], SCOPE [9], POWDER [11], and AERPAW [38]. However, these testbeds are experiment-specific, and in most cases, their geographical restriction prevents them from being globally available.

# 6 Discussion and key takeaways

While the development of 5G technologies has kicked into high gear, there is still a disconnection between the higher-layer CC protocol and the lower layer mmWave PHY/MAC designers. It is crucial to enable higher-layer CC designers to test their implementations on top of realistic lower-layer 5G mmWave-based environments. Therefore, we aim to provide a benchmarking platform *Zeus* to somewhat fill this gap by allowing evaluation and analysis of multiple protocols under different 5G channel conditions and aid in designing new protocols for such settings.

Our analysis of an extensive array of existing protocols evinces a significant gap between most protocol performances and channel capacity in 5G networks. The question of which is the best protocol for such environments remains open. Some key takeaways from our analysis are: (i) Classic loss-based congestion control like Cubic cannot provide low delays in 5G networks. (ii) Although Google's BBR has a more acceptable solo performance in 5G environments with infinite buffers than other CCs; it does not provide compatible behavior alongside TCP-Cubic in different queue sizes. BBR's aggressive behavior in shallow buffers completely dominates Cubic, whereas its generous behavior in deep buffers allows Cubic to dominate. Although they may co-exist fairly in a particular buffer setting, choosing appropriate buffer sizes for their fair operation may not be feasible, especially in 5G cellular networks with over-provisioned buffers. (iii) Copa is a stable protocol and strictly maintains delays at a specific limit. However, it compromises throughput in the process. Copa uses (Standing RTT - minimum RTT) to control its sending rate. The intuition behind Copa's Standing RTT idea is that if the sending rate is less than the link capacity, the queuing delay would be zero at least once every RTT. This behavior might be suitable for delay-sensitive applications like real-time voice and video; however, it is not ideal to support high bandwidth-sensitive applications. (iv) Allegro's fixed-step to modify its sending rate seems to jeopardize its performance in 5G channel capacity utilization. (v) Verus and Vivace bring unstable performance in experiments over the same trace, and do not show good results when co-existing with the default Cubic, (vii) Online learning algorithms, like Vivace and PCC-Proteus, also show highly variable performances, making them unpredictable in 5G networks.

In summary, an essential gap in congestion control research in the 5G domain still exists. The research community must evaluate CC schemes for 5G over broader network conditions and from several aspects, such as; protocols' stability over drastically changing 5G network conditions, efficient behavior over a range of buffer sizes, and consistent behavior alongside existing protocols, particularly from the perspective of the harms they cause to each other. CC algorithms must satisfy the different service requirements to co-exist without negatively affecting each other to rise to the 5G challenge.

## A  Traces generation methodology

### A.1  WiGig router (60 GHz)

These traces were all collected in an indoor office environment. The WiGig router was connected over Ethernet to a Linux machine, and over the WiGig link to an Acer TravelMate P648 laptop, which has a WiGig-enabled network card. A UDP sender at the Linux machine sent packets of fixed size (1500 bytes) over UDP at line rate to the laptop, and timestamp of each received packet was recorded on the laptop. This log was used to compute the inter-arrival times between the packets, which were then used to create a channel trace file that contains the number of arrivals in every unit of time (millisecond) for the entire duration of the trace. When emulating this channel link, the number of arrivals would correspond to the available bandwidth in that millisecond. We collected four WiGig channel traces – *fan_running*: The laptop and the router are placed on the same desk, with a running fan placed in between them. The fan causes slight disturbances in the link; *human_motion*: Random human motion is introduced between the laptop and router placed a few feet apart from each other in an office. The channel capacity is greatly affected by signal attenuation due to the presence of a human body; *stationary*: The laptop is placed next to the router at line of sight. This trace alone was collected in a different indoor environment; *walk_and_turn*: A person holding the laptop moves a few meters towards the router and then turns around.

### A.2  NS-3 scenarios

Although *Zeus* offers a tool to record real-world channel traces, this is an expensive approach. Therefore, *Zeus* also provides a customized NS-3 tool that enables the generation of 5G channel traces for different reference 5G mmWave scenarios. This allows us to define scenarios with desired characteristics aligned with the properties of CC solutions of interest, such as users' motion, capacity sharing or location (i.e., urban, rural). It also defines detailed communication configuration, including modulations and the number or type of antennas. Although NS-3 provides high flexibility in the generation of scenarios (written in C++), it requires in many cases a deep knowledge of the simulator structure and implemented models. We have defined reference scenarios with different characteristics and simplified utilities to customize its operation to overcome this. These utilities include cellular stack settings (including buffer lengths, scheduling, and antenna configuration), users' mobility configuration (predefined track, static, constant speed, and random walks), and generation of performance traces at different layers. In addition, each scenario can be executed with a different random seed to obtain various traces with the same characteristics.

We have divided the NS-3 scenarios into two groups: (i) with buildings and predefined users' motion, and (ii) without buildings and with random users' motion. First, the two most straightforward scenarios (`Buildings`$_1$ and `Buildings`$_2$) comprise one user walking along a street between two buildings towards and away from a base station. Then, in the `Buildings`$_3$ scenario we increment the complexity of the setup by deploying a grid of 3 × 3 buildings and defining a different users track. The `StreetCanyon` scenario comprises two base stations, and the users move following a straight line getting close or far to each of the access elements. Unlike the previous configurations, in the latter the buildings are randomly deployed to obtain different topologies in this setup. All scenarios with buildings utilizes the urban macro (UMa) 3GPP propagation models, except the `StreetCanyon` scenario where the urban micro (UMi) *street canyon* 3GPP model is used. The second group of scenarios represents open areas with different characteristics. In particular, we exploit the UMa, rural macro (RMa) and office indoor hotspot (InH) mixed office 3GPP models to generate three scenarios with the same names (namely, `UMa`, `RMa`, `InhOfficeMixed`) [68]. Unlike the building scenarios, where the user mobility is predefined, in this case, users move following a random pattern with different average distances.

In all of the above cases, the traffic is sent from a server to the user with constant bit-rate, and a packet length of 1472 `bytes`. After the simulation, we obtain a trace that logs all the reception events at the application layer. Although the main objective of this module is the generation of the transport layer trace, the simulation of the scenarios also generates traces from lower layers. Among others, we have implemented a tailored trace generation for the link layer to record the RLC buffer's evolution, which can be seen as the bottleneck buffer when the radio conditions are poor.

Finally, to simplify the generation of traces and their analysis, we have developed a set of `Python` scripts to automate the execution of scenarios varying the data rate, simulation time, and random seed. As a result, we obtain separated trace files for each scenario and configuration. The scripts also use the generated traces to automatically create plots that show the temporal evolution of some parameters (i.e., communications capacity, SINR, RLC, buffer) and the scenario topology (i.e., buildings, access elements, and users position). As has been commented, we use UDP as a transport protocol in saturation. It permits the generation of traces that show the maximum achievable capacity.

In all cases, it is worth noting that *Zeus* is designed to be an extensible framework. Therefore, it contains client/server software and simulation tools to generate traces with different technologies and scenario characteristics.

## B  Zeus' operation validation

**Zeus' operation validation**
We validate the correct operation of *Zeus* by comparing its performance with that obtained in a real environment. In particular, in Figure 16 we represent the statistical distribution of the delay and throughput with whisker plots when using *BBR* as CC solution. The delay is measured as the time elapsed since a packet leaves the sender until an Acknowledgement (Ack) is received back at the sender, confirming the packet delivery. In contrast, the throughput is periodically measured every second, taking the number of bits received within the last second. In Figure 16 we show the results obtained from 3 5G cellular connections, each lasting 60 seconds, using *BBR* during the connections, and results obtained by *Zeus* for *BBR* over similar channel traces recorded at the exact location where the 3 5G cellular connections are created.

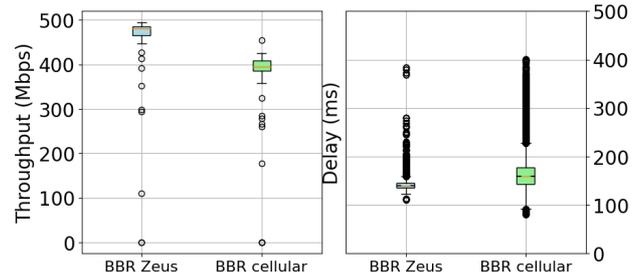

Figure 16: Throughput and delay comparison of *BBR* in a real 5G cellular connection Vs. *Zeus*

In Figure 16 we can see that the results obtained with *Zeus* are similar to those obtained from the actual 5G connections, both in terms of average value and sparsity. In particular, the throughput obtained in both cases, *Zeus* and cellular, reach comparable average values and tight distribution. In the case of the delay, the average values are again alike, while the sparsity observed for the *BBR* cellular connection is slightly larger. Although the results are not identical, they serve to ensure that the CC protocols experience similar and comparable conditions with *Zeus* and an actual cellular connection. Obtaining identical results would not be possible, since the wireless channel realizations are different for the 5G connection running *BBR* and those using UDP to generate the traces used by *Zeus*. The reader may refer to Appendix A for further explanation about the methodology to generate traces from real 5G cellular measurements.

## C  Self-harm of the CC protocols

This Annex shows the results related to the harm that a new flow of a CC protocol may have over existing flows of the same type.

First, Figure 17 depicts the self-harm of the more performing CC protocols, upon different buffer sizes, in the real 5G cellular channel trace. It is measured by comparing the performance of a single flow of each CC protocol with its performance when it shares the channel with another flow of the same type. We do not include Cubic's results since it has been already studied in detail in Figures 9, 10 and 11. In the upper figures, Figure 17a, 17b and 17c, we show the throughput harm and indicate with a dotted line the 50%, which would represent the case where both flows having same throughput. As can be seen, BBR exhibits a harm value slightly below

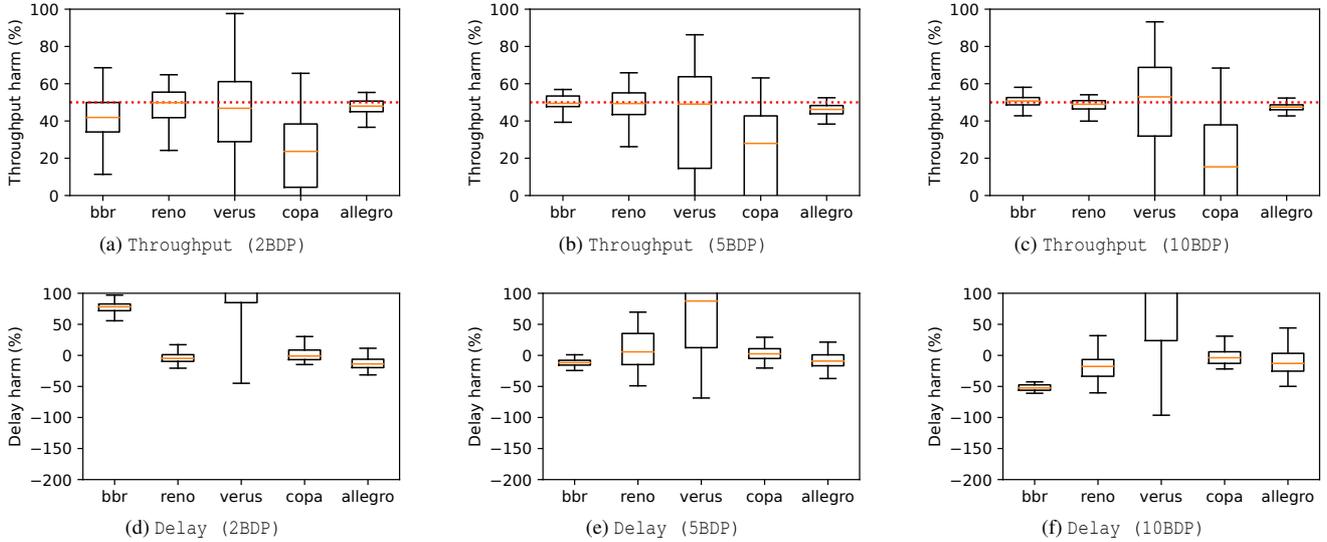

Figure 17: Effect of buffer size on the CC protocols' self-harm for the 5G city driving channel trace.

50 % for shallow buffers, indicating that the protocol performance is not much affected in that case. As we increment the BDP buffer size, BBR's throughput harm tends to 50%. In the case of Reno and Allegro, the results show that the throughput harm is very close to 50 % across the various bottleneck buffer size. Similarly, Verus always presents median throughput self-harm close to 50 %, but with significant variance, evincing its unpredictable behavior. On the other hand, Copa's self-harm is always below 50 % and relatively stable regardless of the bottleneck buffer size. Although Copa's throughput self-harm also has a large variance, the results show its robustness upon flows of the same type. It is worth noting that negative harm values indicate improved delays when sharing the bottleneck with the competing flows. In fact, these unpredictable results yielded by Verus and Copa would indicate some lack of adaptability to the channel.

Figures 17d, 17e and 17f show the delay self-harm. BBR's delay self-harm strongly depends on buffer size, going from around 75% (negative impact) for shallow buffers, to -50% (positive impact) in deeper ones. On the other hand, Reno, Copa and Allegro present a rather stable delay self-harm which is, in average, around 0 (no impact) for the different buffer sizes. Finally, we can see that Verus presents a high variance in this metric, as was already observed for the throughput self-harm.

To complement the previous results, in Figure 18 we plot the temporal evolution of the throughput and delay when a flow runs alone and when two flows share the bottleneck buffer. In all cases, the flow alone and the competing flows are represented with and without sub-indices, respectively. For instance, in Figure 17a the legend entry *BBR* refers to the performance of a BBR flow alone, and the legend entries *BBR1* and *BBR2* to the performance of the competing flows sharing the bottleneck buffer.

The results are aligned to the previous ones. BBR, Reno and Allegro, Figures 18a-18c, 18d-18f and 18m-18o respectively, show a good sharing of the channel capacity. As can be observed, the throughput with the competing flows is around half of that seen with a flow alone, and more importantly the competing flows have very similar performance. Also, in the case of Copa, the competing flows have similar behavior. However, we see in Figure 17 that the throughput harm was, in general, below 50%. In Figures 18j-18l, we can observe that Copa with a single flow is far from fully achieving the channel capacity so that the competing flows can achieve a throughput above the middle of that reached by a single flow. It is worth noting that other metrics, like Jain's fairness index, would indicate a good sharing of the bottleneck buffer. As expected, Verus results shown in Figure 18g-18i reflect the unpredictable behavior of the protocol, where the performance of the competing flows is above that yielded by the flow alone. As can be observed in, for instance, Figure 18h, in the time slot between 10 and 20 seconds, the throughput reached by the flow alone was much below the channel capacity. However, when competing flows are deployed, one of them almost reaches the channel capacity, which would lead to improvement and so negative harm. In sum, the harm Verus' values shown in Figure 17 again reflects the lack of adaptation to the fluctuations of the channel capacity.

## D Ethics

This work does not raise any ethical issues.

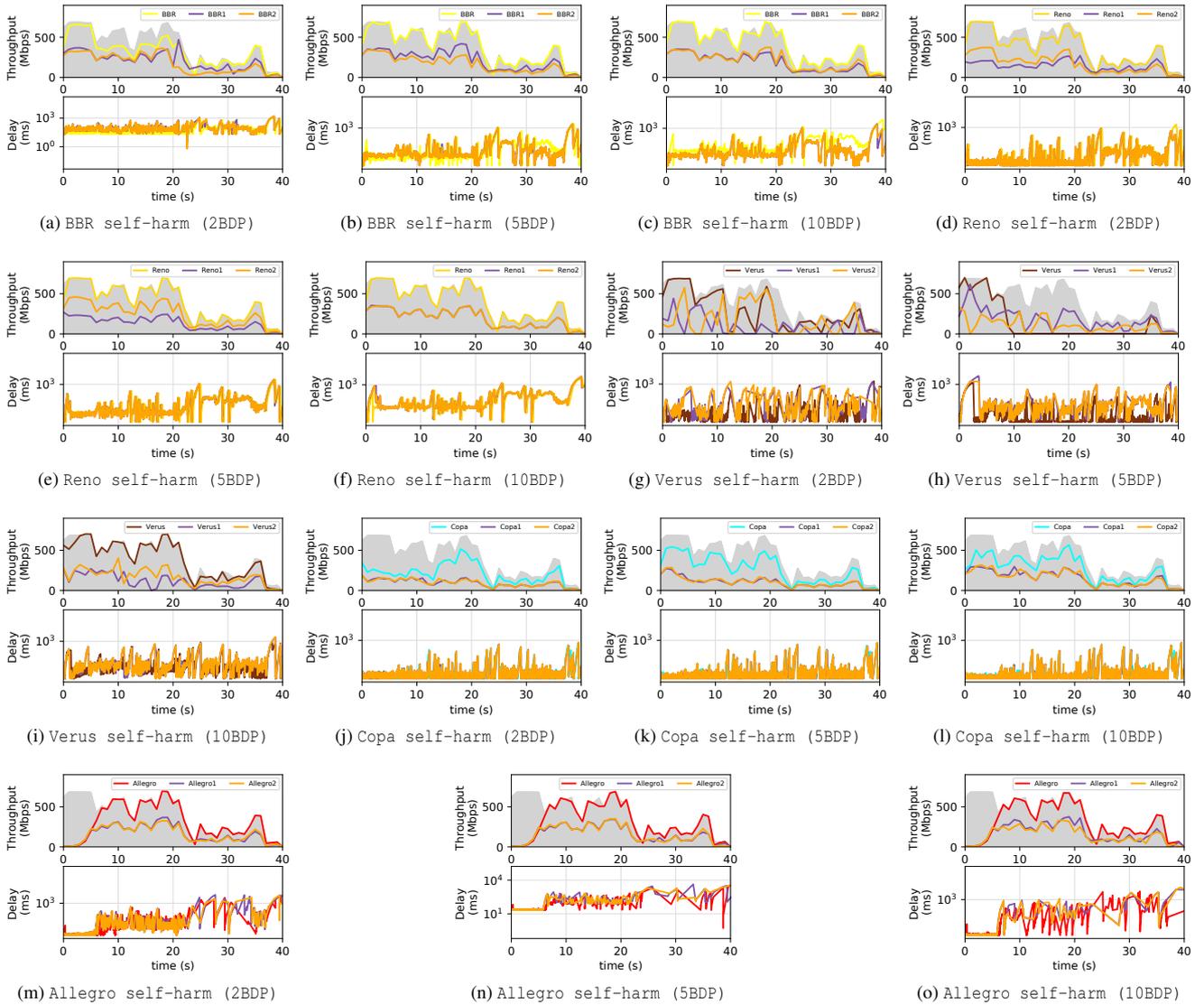

Figure 18: Effect of buffer size on the CC protocols' self-harm for the 5G city driving channel trace.